\documentclass{PoS}

\usepackage{graphicx}

\newcommand{\be}{\begin{equation}}
\newcommand{\ee}{\end{equation}}
\newcommand{\beq}{\begin{eqnarray}}
\newcommand{\eeq}{\end{eqnarray}}

\PoS{PoS(LAT2005)091}

\title{A study of the $N$ to $\Delta$ transition form factors in full QCD }

\ShortTitle{A study of the $N$ to $\Delta$ transition form factors in full QCD}

\author{Constantia Alexandrou \\
Department of Physics, University of Cyprus, CY-1678 Nicosia, Cyprus\\
E-mail: \email{alexand@ucy.ac.cy}}
\author{Robert Edwards\\
Thomas Jefferson National Accelerator Facility, 
                   Newport News, VA 23606, USA \\
E-mail: \email{edwards@jlab.org}}

\author{Giannis Koutsou \\
Department of Physics, University of Cyprus,
                   CY-1678 Nicosia, Cyprus \\
E-mail:\email{koutsou@ucy.ac.cy}}
\author{ Theodoros Leontiou \\
Department of Physics, University of Cyprus,
                   CY-1678 Nicosia, Cyprus \\
E-mail:\email{t.leontiou@ucy.ac.cy}}
\author{Hartmut Neff \\
Centre for Computational Science, Chemistry Department,
                   University College of London, \\
                   20 Gordon Street, WC1H 0AJ, London, UK \\
E-mail: \email{ucchane@ucl.ac.uk}}
\author{John W. Negele \\
Center for Theoretical Physics, Laboratory for Nuclear 
                   Science and Department of Physics, Massachusetts Institute
                   of Technology, Cambridge MA 02139, USA \\
E-mail:\email{negele@lns.mit.edu}}
\author{Wolfram Schroers \\
NIC/DESY Zeuthen, Platanenallee 6, D-15738 Zeuthen, 
                   Germany \\
E-mail:\email{Wolfram.Schroers@Feldtheorie.de}}
\author{\speaker{Antonios Tsapalis} \\
        University of Athens, Institute of Accelerating Systems
	and Applications, Athens, Greece \\        
        E-mail: \email{tsapalis@cc.uoa.gr}}

\abstract{The $N$ to $\Delta$ transition form factors $G_{M1}$, $G_{E2}$ 
and $G_{C2}$ are
evaluated using dynamical MILC configurations and valence domain wall fermions
at three values of quark mass corresponding
to pion mass 606 MeV, 502 MeV and 364 MeV 
on lattices of spatial size $20^3$ and $28^3$.
The unquenched results are compared to those obtained at similar pion mass
in the quenched theory.}

\FullConference{XXIIIrd International Symposium on Lattice Field Theory\\

                 25-30 July 2005\\

                 Trinity College, Dublin, Ireland}

\begin{document}

\section{Introduction}
Non-zero quadrupole strength in the N to $\Delta$  transition is interpreted
 as a
signal for deformation in the nucleon and/or $\Delta$. 
Hadron deformation  is a fundamental  property 
   that depends on QCD dynamics and
determining the shape of the nucleon has motivated experimental measurements
 of quadrupole form factors in the N to $\Delta$ transition for
many years. 
Selection rules in the N to $\Delta$ transition
allow for a magnetic dipole, M1,
an electric quadrupole, E2, and a Coulomb quadrupole, C2, 
with the corresponding transition form factors $G_{M1}$, 
$G_{E2}$ and $G_{C2}$ as functions of the
four-momentum transfer squared, $q^2$.  
The magnetic transition, which proceeds through a quark spin-flip,
is expected to be the dominant one while non-zero quadrupole strengths can
be obtained in many models by assuming a deformed 
nucleon  and/or $\Delta$ as for example in the quark model 
where a D-wave admixture is assumed in the nucleon and $\Delta$ wave functions.
Pion electro- and photo-production 
experiments near the $\Delta (1232)$ resonance have
 clearly established non-zero
quadrupole strength in the range of a few percent  with respect to the
dominant M1 strength providing
support for deformation~\cite{costas}. 
In ref.~\cite{bates}, a precise measurement at
$-q^2 = 0.127 \> {\rm GeV}^2$ extracted the values
\be
{\rm R_{EM}} \equiv -\frac{\rm G_{E2}}{\rm G_{M1}} = (-2.3 \pm 0.3 
\pm 0.6 )\> \% , \hspace*{0.5cm} 
{\rm R_{SM}} \equiv-\frac{|{\bf q}|}{2m_\Delta} \frac{\rm G_{C2}}{\rm G_{M1}} = (-6.1  \pm 0.2
\pm 0.5) \> \% 
\label{ratios}
\ee 
for the ratio of the electric and coulomb quadrupole to the 
magnetic dipole form factors,
 $R_{EM}$ (or EMR) and $R_{SM}$ (or CMR).
The first uncertainty given in Eq.~(\ref{ratios}) 
is the  statistical and systematic error
and the second the model error. 
At the same time experiments at Jefferson Lab
have studied the $q^2$ dependence of these three form factors 
in the range of $0.4 - 1.8 \> {\rm GeV}^2$~\cite{jlab}.

Within lattice QCD one can evaluate 
these transition form factors starting directly from
the QCD Lagrangian.
 The first lattice study of the  $N$ to $\Delta$ transition 
was carried out very early on~\cite{leinweber}
and, although the quadrupole form factors were not determined to
high enough accuracy, the methodology for the calculation of
 this matrix element
on the lattice
was presented. 
 Using the formalism laid out in Ref.~\cite{leinweber}
a more detailed study followed in Ref.~\cite{prd}
which established the 
existence of small and negative electric quadrupole strength 
for a given momentum transfer ${\bf q}$  within quenched QCD.
In subsequent work~\cite{lat04,prl} we focus
 on the determination of the momentum 
dependence of the transition form factors. We carry out a number
of improvements that enables us to extract both
electric and coulomb quadrupole form factors to high enough
accuracy to exclude a zero value. The improvements come from
optimizing the interpolating field used for the $\Delta$
 so that the three-point function
is evaluated for the  maximum allowed values of lattice momentum vectors for
a given $q^2$
 and from using a simultaneous overconstrained analysis of the 
resulting matrix elements. We obtain
small negative values for the ratios $R_{EM}$ and $R_{SM}$  
in qualitative agreement with the experimental measurements
for values of $-q^2$ in the range $0.4$ to $1.5$ GeV$^2$. 
On  the other hand, the values obtained for $R_{SM}$
are smaller than experiment for $-q^2<0.4$~GeV$^2$.
Similarly
the values obtained for  $G_{M1}$ in this quenched study, after
linearly extrapolated to the chiral limit, disagree 
 with the experimental measurements raising questions about the
importance of sea quark effects and the validity of the linear
extrapolation of the lattice data.
 Recently a relativistic chiral
effective-field theory calculation of the $R_{EM}$ and $R_{SM}$
ratios~\cite{chieff}
shows that the linear extrapolation
is incorrect at low $q^2$ and a stronger dependence on the pion mass 
sets in at a pion mass value where the $\Delta$ becomes unstable.
This stronger mass dependence drives $R_{SM}$ to more
negative values resolving the discrepancy between lattice
results and experiment at low $q^2$. This study explicitly demonstrates
the importance of pion cloud contributions in the determination
of these transition form factors.

In this work we attempt to study the effects
of unquenching 
on the three transition form factors 
using  dynamical MILC configurations for the sea quarks
and employing
domain wall valence fermions. Given the fact that chiral
fermions are still very expensive to simulate
 this ``hybrid'' scheme appears as a reasonable compromise and has already
been employed in heavy quark spectroscopy and generalized parton
distributions calculations \cite{lhpc}.

\section{Lattice techniques}
The matrix element for the $N$ to $\Delta$ electromagnetic transition 
for on-shell nucleon and $\Delta$ states and real or virtual photons has
the form~\cite{jones} 
\beq
 \langle \; \Delta (p',s') \; | j_\mu | \; N (p,s) \rangle =    
 \> i \sqrt{\frac{2}{3}} \biggl(\frac{m_{\Delta}\; m_N}{E_{\Delta}({\bf p}^\prime)\;E_N({\bf p})}\biggr)^{1/2} 
  \bar{u}_\tau (p',s') {\cal O}^{\tau \mu} u(p,s) \; 
\eeq
where $p,s\>$ and $p',s'\>$ denote initial and final momenta and spins and 
$ u_\tau (p',s')$ is a spin-vector in the Rarita-Schwinger formalism.
The operator 
${\cal O}^{\tau \mu}$  can be decomposed in terms of the Sachs form factors
as
\be
{\cal O}^{\tau \mu} =
  {G}_{M1}(q^2) K^{\tau \mu}_{M1} 
+{G}_{E2}(q^2) K^{\tau \mu}_{E2} 
+{G}_{C2}(q^2) K^{\tau \mu}_{C2} \;,
\ee
where the exact expressions for the kinematical functions $K^{\tau \mu}$ can be found in ref.~\cite{prd}.
%In the rest frame of the $\Delta$, the ${\rm R_{EM}}$ and ${\rm R_{SM}}$ 
%ratios (also known as EMR and CMR respectively) are obtained via
%\be
% R_{EM}= -\frac{{G}_{E2}(q^2)}{{G}_{M1}(q^2)} \>,  
%\hspace*{0.8cm}
% R_{SM}=-\frac{|{\bf q}|}{2m_\Delta}\;\frac{{G}_{C2}(q^2)}{{G}_{M1}(q^2)} 
%\quad.
%\ee
The extraction of the Sachs form factors requires the computation of the
three-point function $G^{\Delta j^\mu N}_{\sigma} (t_2,t_1 ; {\bf p}^{\;\prime}, {\bf p};\Gamma )$ along with the nucleon and $\Delta$ two-point functions
$ G^{NN}$ and $ G^{\Delta \Delta}_{ij}$. The nucleon source is
taken at  time zero, 
the photon is absorbed by a quark at a later time $t_1$ and the $\Delta$ sink
is at an even later time $t_2$.
Provided the Euclidean time separations $t_1$ and  $t_2-t_1$
 are large enough, the
time dependence and field renormalization constants will cancel in the ratio
\small
\beq
& & R_\sigma (t_2,t_1; {\bf p}^{\; \prime}, {\bf p}\; ; \Gamma ; \mu) =
\frac{\langle G^{\Delta j^\mu N}_{\sigma} (t_2, t_1 ; {\bf p}^{\;\prime}, {\bf p};\Gamma ) \rangle \;}{\langle G^{\Delta \Delta}_{ii} (t_2, {\bf p}^{\;\prime};\Gamma_4 ) \rangle \;} \> \times \nonumber \\
 &\>&  \biggl [ \frac{ \langle G^{N N}(t_2-t_1, {\bf p};\Gamma_4 ) \rangle \;\langle 
G^{\Delta \Delta}_{ii} (t_1, {\bf p}^{\;\prime};\Gamma_4 ) \rangle \;\langle 
G^{\Delta \Delta}_{ii} (t_2, {\bf p}^{\;\prime};\Gamma_4 ) \rangle \;}
{\langle G^{\Delta \Delta}_{ii} (t_2-t_1, {\bf p}^{\;\prime};\Gamma_4 ) \rangle \;\langle 
G^{N N} (t_1, {\bf p};\Gamma_4 ) \rangle \;\langle 
G^{N N} (t_2, {\bf p};\Gamma_4 ) \rangle \;} \biggr ]^{1/2} 
 \stackrel{t_2 -t_1 \gg 1 ,\; t_1 \gg 1}{\longrightarrow}
\Pi_{\sigma}({\bf p}^{\; \prime}, {\bf p}\; ; \Gamma ; \mu) \>,
\label{R-ratio}
\eeq
\normalsize
where $\sigma$ denotes the vector index of the $\Delta$ field. The
matrices $\Gamma$ 
are projections  for the Dirac indices
\be
\Gamma_i = \frac{1}{2}
\left(\begin{array}{cc} \sigma_i & 0 \\ 0 & 0 \end{array}
\right) \;\;, \;\;\;\;
\Gamma_4 = \frac{1}{2}
\left(\begin{array}{cc} I & 0 \\ 0 & 0 \end{array}
\right) \;\;. 
\ee
We fix the $\Delta$ at rest and therefore ${\bf q}={\bf p}^{\prime}-{\bf p}
=-{\bf p}$. $Q^2=-q^2$ is the Euclidean momentum transfer squared.
Determining $\Pi_{\sigma}({\bf q}\; ; \Gamma ;\mu)$ for
given values of $\sigma$ and $\Gamma$ by fitting to the
plateau of $R_\sigma (t_2,t_1; {\bf p}^{\; \prime}, {\bf p}\; ; \Gamma ; \mu)$
enables us to obtain 
the Sachs form factors. For example,
\be 
\Pi_{\sigma}({\bf q}\; ; \Gamma_4 ;\mu)= i A \epsilon^{\sigma 4\mu j} p^j 
{G}_{M1}(Q^2) \quad ,
\label{pure GM1}
\ee
with A a kinematical coefficient. Other similar relations can be found in 
Ref.~\cite{lat04}.
A novel feature of Ref.~\cite{prl} is the choice 
of linear combination of  ratios 
$R_\sigma (t_2,t_1; {\bf p}^{\; \prime}, {\bf p}\; ; \Gamma ; \mu)$ 
for appropriate values of $(\sigma,\Gamma)$
such that, for given
$Q^2$, a maximum number of photon momentum transfers ${\bf q}$ give 
non-vanishing contributions. The optimal combination for $G_{M1}$, referred to
as $S_1$-type, is
\be
S_1({\bf q};\mu)=  \sum_{\sigma=1}^3\Pi_\sigma({\bf q}\; ; \Gamma_4 ;\mu) =
i A \left[ (p_2-p_3)\delta_{1,\mu}
+ (p_3-p_1)\delta_{2,\mu} +  (p_1-p_2)\delta_{3,\mu}\right ] G_{M1}(Q^2)
\label{S1}
\ee
where all the vectors ${\bf q}$ related through lattice rotations participate.
Similarly, the quadrupole form factors are obtained from type $S_2$ given by 
\beq
S_2({\bf q};\mu) &=&\sum_{\sigma\neq k=1}^{3} \Pi_\sigma({\bf q}\; ; 
 \Gamma_k ;\mu)  \nonumber \\  &=&  
B \Biggl\{  \left[ (p_2+p_3)\delta_{1,\mu}
+ (p_3+p_1)\delta_{2,\mu} +  (p_1+p_2)\delta_{3,\mu}\right ] G_{E2}(Q^2)
\nonumber \\
&-& 2 \frac{p_\mu}{{\bf p}^2}(p_1 p_2 + p_1 p_3 + p_2 p_3) \left[ G_{E2}(Q^2)
+ \frac{E_N-m_\Delta}{2 m_\Delta} G_{C2}(Q^2) \right]
\Biggr\} \; ,
\label{S2}
\eeq
for   $\mu = 1,2,3 $. 
For $\mu = 4 \>$ we have 
\be
S_2({\bf q};\mu = 4)= \frac{6\;C}{{\bf p}^2} (p_1 p_2 + p_1 p_3 + p_2 p_3) 
G_{C2}(Q^2) \> ,
\ee
where $B$ and $C$ are kinematical factors.
The  three-point functions needed for the evaluation
of $S_1$ and $S_2$ are obtained by fixing the appropriate 
combination of interpolating fields for the $\Delta$ and require only 
{\it one sequential } inversion for each fixed sink type. 
Therefore, two independent sequential inversions suffice for the complete
evaluation of the form factors for all available lattice momenta ${\bf q}$. 
The full set of lattice measurements for any available $\mu$ and ${\bf q}$
for a given $Q^2$ are simultaneously analyzed 
and the form factors are extracted from a global $\chi^2$-minimization 
procedure~\cite{prl}.

The dynamical configurations that we are using
are generated by the MILC collaboration and are 
available from the NERSC archive. For the lattice of volume $20^3 \times 64$
we use the ensembles with
the strange quark mass fixed at $am_s=0.05$ and the light quark flavors
at $am_{u,d}=0.03$ and $0.02$ simulated with  the improved 
staggered (Asqtad) action which ensures
better scaling properties at the lattice spacing of $a=0.125\;
{\rm fm}$ used in the simulations. 
In addition we use configurations generated on a lattice
of size $28^3 \times 64$ i.e. 
physical spatial volume of
$(3.5 \; {\rm fm})^3$ at $am_s=0.05$  and $am_{u,d}=0.01$.
We apply hypercubic (HYP) smearing to the configurations and use
Domain wall  
fermions
in the valence sector.  The size of the fifth dimension 
is set to $L_5=16$ where it has been shown \cite{lhpc} that, for 
these lattices, the residual 
quark mass, computed from the divergence of the five-dimensional axial 
current, is at least an order of magnitude smaller than the bare mass.
 The height of the domain wall parameter is set at $am_0 = 1.7$
and  Dirichlet boundary conditions are imposed at $t/a=32$ since
only half the time extent of the lattice is needed.
Smeared quark wave functions are known to increase the overlap of the 
interpolating fields with the hadronic states reducing  the
contamination from excited states. Before the calculation of the
valence propagators we apply 
%30 iterations of 
gauge invariant
Wuppertal smearing
%(with  $\alpha=0.5$ in front of the hopping matrix)
 to the local quark fields $q(x)$ via
 $q_{\rm smear}({\bf x},t)=\sum_{\bf y}(1+\alpha H)^n({\bf x},{\bf y}) q({\bf y},t)$ with $H$ the hopping matrix, 
$\alpha\sim 3$
and $n=30$.

 The bare Domain wall quark mass parameter $(am_q)^{DW}$ has been tuned by 
the LHP collaboration \cite{lhpc} 
by requiring that the pion mass computed with the Domain wall valence 
quarks equals the mass of the lightest pion
computed with the staggered fermions.
The resulting quark mass parameters and corresponding pion masses obtained
by the LHP collaboration are shown in
Table 1. In the same Table we also give our values for
 the nucleon and $\Delta$ masses. As can be seen the mass of the $\Delta$ 
carries the largest errors.

\begin{table}[ht]
\begin{center}
\label{table:par}
\begin{tabular}{c c c c c c c c}
\hline
Volume  &  $(am_{u,d})^{\rm sea}$ & $(am_s)^{\rm sea}$ &
$(am_q)^{DW}$ & $m_{\pi}^{DW}$ (GeV) & $ m_{\pi}/m_{\rho}$ &$m_N$~(GeV) &$m_\Delta$~(GeV)\\ 
\hline
\hline
$20^3 \times 32$   & 0.03 & 0.05 & 0.0478 & 0.606(2) & 0.588(7) & 1.392(9) & 1.662(21) \\
\hline
$20^3 \times 32$   & 0.02 & 0.05 & 0.0313 & 0.502(4) & $\>\>\;$0.530(11) &1.255(19) & 1.586 (36)\\
\hline
$28^3 \times 32$   & 0.01 & 0.05 & 0.0138 & 0.364(1) & 0.387(7) & 1.196(25) & 1.643(63) \\
\hline
\end{tabular}
\caption{Mass parameters for the sea (Asqtad) and valence (DW) quarks and 
corresponding meson masses taken from \cite{lhpc}. In the last
two columns we give our values for the nucleon and $\Delta$ mass
using 125 configurations for $am_{u,d}=0.03$, 75 for  $am_{u,d}=0.02$ and
38 for  $am_{u,d}=0.01$.}
\end{center}
\end{table}
\vspace*{-1.cm}

\section{Results and outlook}

\begin{figure}[ht]
\vspace*{-0.5cm}
\begin{minipage}{6.5cm}
\hspace*{-0.8cm}
{\mbox{\includegraphics[height=6.2cm,width=8.3cm]{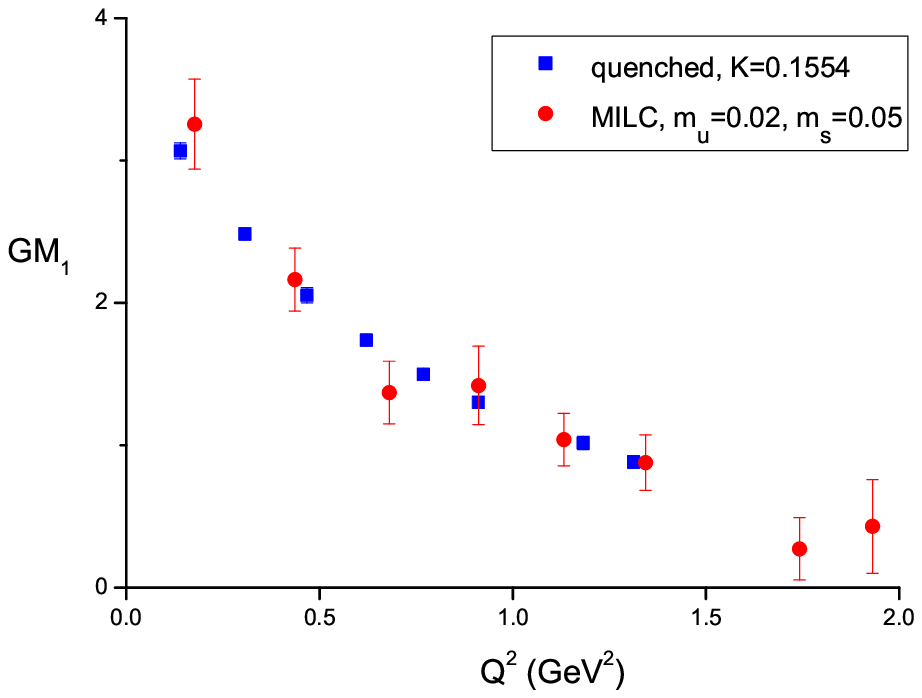}}}
\vspace*{-0.5cm}
\caption{$G_{M1}$ in units of $(e/2m_N)$ in quenched and $N_f=2+1$ QCD at $m_{\pi}\sim 0.5$ GeV.}
\label{fig:GM1compare}
\end{minipage}
\hspace{0.7cm}
\begin{minipage}{7cm}
{\mbox{\includegraphics[height=6.2cm,width=8.3cm]{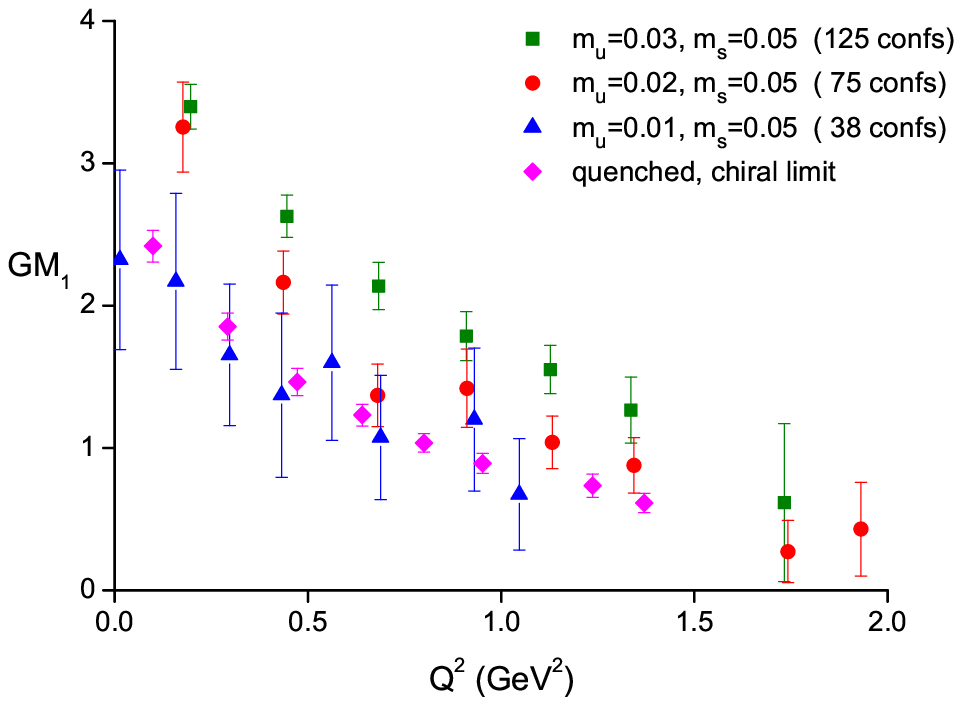}}}
\vspace*{-0.5cm}
\caption{$G_{M1}$ in units of $(e/2m_N)$ using MILC configurations 
for three quark masses and quenched results at the chiral limit.}
\label{fig:GM1}
\end{minipage}

\end{figure}

\begin{figure}[ht]
\vspace*{-0.5cm}
\begin{minipage}{6.5cm}
\hspace{-0.5cm}
{\mbox{\includegraphics[height=6.2cm,width=8.3cm]{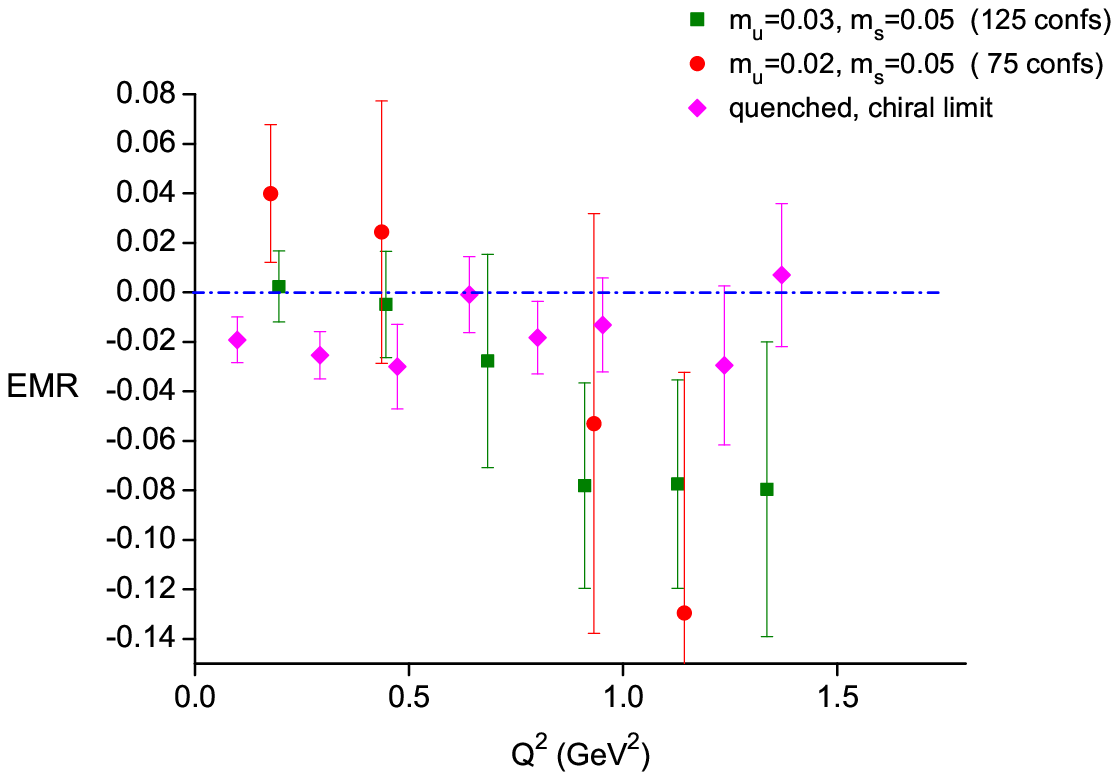}}}
\vspace*{-0.8cm}
\caption{EMR ratio using MILC configurations at $am_{u,d}=0.03, 0.02$
 and quenched results at the chiral limit.}
\label{fig:EMR}
\end{minipage}
\hspace{0.5cm}
\begin{minipage}{6.5cm}
{\mbox{\includegraphics[height=6.2cm,width=8.3cm]{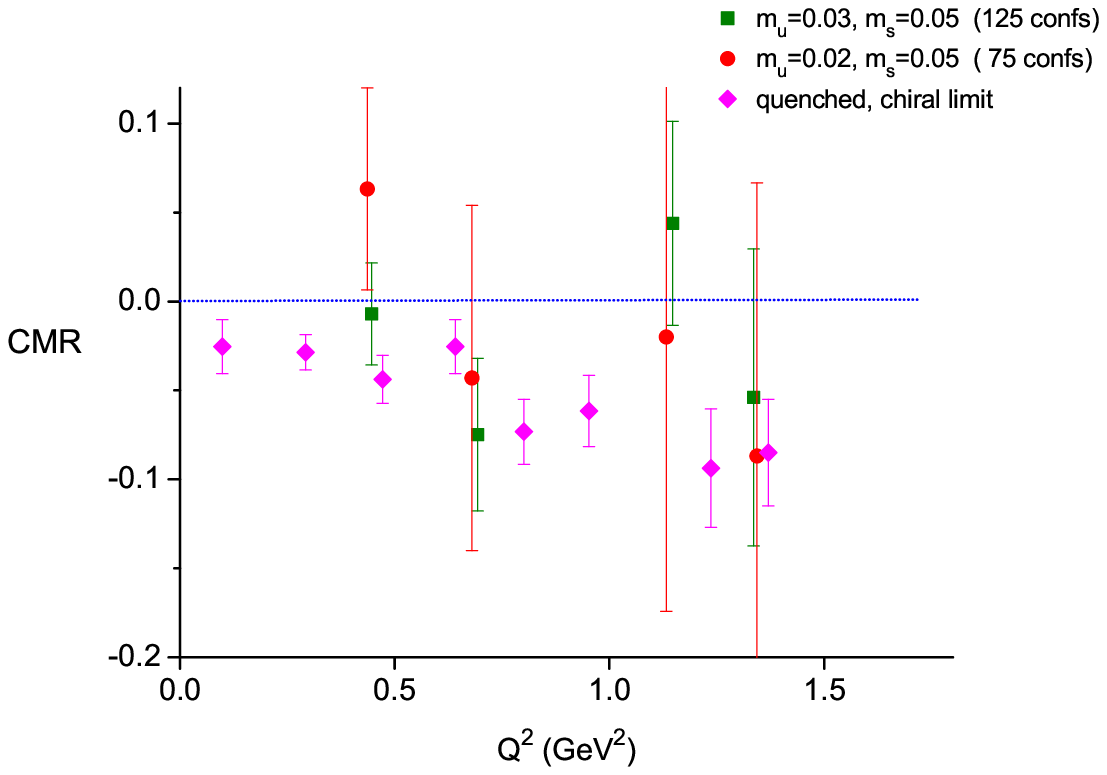}}}
\vspace*{-0.8cm}
\caption{CMR ratio using MILC configurations at $am_{u,d}=0.03, 0.02$
 and quenched results at the chiral limit.}
\label{fig:CMR}
\end{minipage}

\end{figure}

For the $20^3 \times 32$ volume we analyzed 125 configurations at
$am_{u,d}=0.03$ and 75 at  $am_{u,d}=0.02$  
using both $S_1$ and $S_2$ types of interpolating fields. This
enable us to extract the dipole and quadrupole form factors. 
In addition, we analyzed 38 configurations on the $28^3 \times 32$ 
at $am_{u,d}=0.01$  
lattice using just the  $S_1$-type, providing a first indication
of the magnetic dipole form factor behavior at a lighter pion mass. 
We employ the local current operator $\bar{\psi}(x)\gamma_{\mu}\psi(x)$, 
which is not conserved and therefore we need 
to multiply by the renormalization factor $Z_V$. This is evaluated
non-perturbatively  by evaluating the elastic nucleon electric
 form factor at $Q^2=0$.

 In Fig.~\ref{fig:GM1compare} we make a comparison of the magnetic dipole 
form factor in quenched and dynamical lattice QCD at a similar pion mass
of about $0.5$ GeV. No unquenching effects can be observed at this rather
heavy pion mass, given the size of the statistical
errors  on the dynamical QCD 
results.
 The results on $G_{M1}$ on the available lattices are presented in 
Fig.~\ref{fig:GM1} along with the  quenched results obtained
by linearly extrapolating to the chiral limit~\cite{prd}. The accuracy
of the results in full QCD do not allow for a chiral extrapolation
but the general trend is that they reach
 lower values as compared to the quenched case,
a welcome fact given  
that the quenched results were higher than experiment~\cite{prl}.
On the other hand, the EMR and CMR ratios calculated on these ensembles 
and presented in Figs.~\ref{fig:EMR} and \ref{fig:CMR}
are still very  noisy 
prohibiting a meaningful comparison to the quenched results.

In conclusion, preliminary results are presented for
 the N to $\Delta$
magnetic dipole transition form factor $G_{M1}$ and the EMR and CMR ratios,
evaluated using Domain wall valence quarks and
dynamical MILC configurations.
The statistical fluctuations are enhanced in full QCD requiring a larger number
of configurations in order to reach the accuracy required for  a
meaningful comparison to either the quenched data or experiment.\\

\vspace*{-0.3cm}

\noindent
{\bf Acknowledgments:}
A.T. acknowledges support from
the program ``Pythagoras'' of the Greek Ministry of Education.
This work has been supported in part by the EU Integrated Infrastructure 
Initiative Hadron Physics (I3HP) under contract RII3-CT-2004-506078.

\end{document}